\RequirePackage{fixltx2e}
\documentclass[aps,prb,reprint,floatfix]{revtex4-1}

\usepackage{amsmath,amsfonts,amssymb} %
\usepackage{mathtools} %
\usepackage{wasysym} %
\usepackage{bm} %
\usepackage[bbgreekl]{mathbbol} %
\usepackage{graphicx} %
\usepackage[dvipsnames,table]{xcolor} %
\usepackage{tabularx} %
\usepackage[acronym]{glossaries} %
\usepackage{braket} %
\usepackage{fancyhdr} %
\usepackage{microtype} %
\usepackage{natmove}
\usepackage[byname]{smartref} %
\usepackage[breaklinks, %
colorlinks, linkcolor=Blue, citecolor=Blue, urlcolor=Blue]{hyperref} %
\usepackage[version=3]{mhchem}
\AtBeginDocument{\usepackage{booktabs}} 

\newcommand{\CR}[1]{a^{\dagger}_{#1}} %
\newcommand{\AN}[1]{a_{#1}} %

\newcommand{\BC}{\color{black}}
\newcommand{\eq}[1]{Eq.~(\ref{#1})} %
\newcommand{\FF}{\mathcal{F}} %
\def\be{\begin{equation}} %
\def\ee{\end{equation}} %
\def\bea{\begin{eqnarray}} %
\def\eea{\end{eqnarray}} %

\newacronym{QPE}{QPE}{quantum phase estimation} %
\newacronym{VQE}{VQE}{variational quantum eigensolver} %
\newacronym{UCC}{UCC}{unitary coupled cluster} %
\newacronym{QCC}{QCC}{qubit coupled cluster} %
\newacronym{FCI}{FCI}{full configurational interaction} %
\newacronym{CASCI}{CASCI}{complete active space configurational
  interaction} %
\newacronym{JW}{JW}{Jordan--Wigner} %
\newacronym{BK}{BK}{Bravyi--Kitaev} %
\newacronym[longplural={degrees of freedom}, %
firstplural={degrees of freedom (DOF)}, plural={DOF}]{DOF}{DOF}{degree
  of freedom} %
\newacronym[longplural={equations of motion}, %
firstplural={equations of motion (EOM)}, %
plural={EOM}]{EOM}{EOM}{equation of motion} %
\newacronym{PES}{PES}{potential energy surface} %
\newacronym{CI}{CI}{configuration interaction} %
\newacronym{QMF}{QMF}{qubit mean-field} %
\newacronym{SQP}{SQP}{sequential quadratic programming} %
\newacronym{RHF}{RHF}{restricted Hartree--Fock}

\begin{document}

\author{Tzu-Ching Yen} 
\affiliation{Department of Physical and Environmental Sciences,
  University of Toronto Scarborough, Toronto, Ontario, M1C 1A4,
  Canada; and Chemical Physics Theory Group, Department of Chemistry,
  University of Toronto, Toronto, Ontario, M5S 3H6, Canada; email of the corresponding author: artur.izmaylov@utoronto.ca}

\author{Robert A. Lang} 
\affiliation{Department of Physical and Environmental Sciences,
  University of Toronto Scarborough, Toronto, Ontario, M1C 1A4,
  Canada; and Chemical Physics Theory Group, Department of Chemistry,
  University of Toronto, Toronto, Ontario, M5S 3H6, Canada; email of the corresponding author: artur.izmaylov@utoronto.ca}

\author{Artur F. Izmaylov}
\email{artur.izmaylov@utoronto.ca}
\affiliation{Department of Physical and Environmental Sciences,
  University of Toronto Scarborough, Toronto, Ontario, M1C 1A4,
  Canada; and Chemical Physics Theory Group, Department of Chemistry,
  University of Toronto, Toronto, Ontario, M5S 3H6, Canada; email of the corresponding author: artur.izmaylov@utoronto.ca}
  
\title{Exact and approximate symmetry projectors for the electronic structure problem on a quantum computer
}

\date{\today}

\begin{abstract}
Solving the electronic structure problem on a universal-gate quantum computer within 
the variational quantum eigensolver (VQE) methodology requires constraining 
the search procedure to a subspace defined by relevant physical symmetries. Ignoring symmetries 
results in convergence to the lowest eigenstate of the Fock space for the second quantized electronic 
Hamiltonian. Moreover, this eigenstate can be symmetry broken due to limitations of the wavefunction 
ansatz. To address this VQE problem, we introduce and assess methods of exact and approximate 
projectors to irreducible eigen-subspaces of available physical symmetries. 
Feasibility of symmetry projectors in the VQE framework is discussed, and their efficiency 
is compared with symmetry constraint optimization procedures. Generally, projectors introduce higher 
numbers of terms for VQE measurement compare to the constraint approach. On the other hand,
the projection formalism improves accuracy of the variational wavefunction ansatz without introducing 
additional unitary transformations, which is beneficial for reducing depths of quantum circuits.       
\end{abstract}

\glsresetall

\maketitle


\section{Introduction}

The \gls{VQE} method\cite{Peruzzo:2014/ncomm/4213, Jarrod:2016/njp/023023, Wecker:2015/pra/042303} 
is one of the most practical approaches to the electronic structure problem 
on current and near future universal quantum computers. 
%
\gls{VQE} finds the best approximation to the lowest eigenstate of a Hamiltonian in a manifold of 
wavefunctions by a variational energy minimization procedure 
involving both quantum and classical computers.
The quantum computer operates in terms 
of qubit operators and qubit wavefunctions; its role is to set up a trial qubit wavefunction and 
to measure its energy on a qubit Hamiltonian for obtaining the energy expectation value.
To reformulate the system electronic Hamiltonian ($\hat H_e$) in a qubit form ($\hat H_q$), 
usually, $\hat H_e$ is taken in the second quantized form and is transformed iso-spectrally using 
the fermion-spin transformations such as  
\gls{JW}~\cite{Jordan:1928/zphys/631, AspuruGuzik:2005/sci/1704} or
more resource-efficient \gls{BK}~\cite{Bravyi:2002/aph/210, Seeley:2012/jcp/224109,
  Tranter:2015/ijqc/1431, Setia:2017/ArXiv/1712.00446,
  Havlicek:2017/pra/032332}.
{\BC In this setup, a quantum hardware measures the expectation value of the energy on the current trial wavefunction. Then, a classical computer determines a new trial wavefunction based on the information collected by the quantum hardware. These two steps are iterated until convergence of the energy.}
The VQE was successfully employed on several quantum computers and 
used for few small molecules up to BeH$_2$\cite{Kandala:2017/nature/242} 
and H$_2$O.\cite{CVQE,IonQ}       

Note that even though the wavefunction optimization problem 
is not solved on a quantum computer, the main advantage of the hybrid VQE scheme 
is a compact representation of a unitary ansatz for the wavefunction. One of the standard 
unitary hierarchies of approximations for the wavefunction is the unitary coupled cluster (UCC) method.
At any finite level of excitations starting from doubles the number of terms in UCC equations 
grow exponentially with the size of the system on a classical computer, while the number of parameters 
for the VQE optimization grows only polynomially. It was found recently that even though the UCC 
hierarchy is generally more efficient than a regular coupled cluster hierarchy in capturing the electron 
correlation energy, it still breaks down for strongly correlated models.\cite{Harsha:2018/jcp/044107}

One of the VQE issues is that, in most implementations, it operates in the Fock space of the original fermionic problem, which maps iso-spectrally onto the Hilbert space of $N$ qubits.
Thus, states of all possible number of electrons 
are present in the qubit Hilbert space. To optimize a particular electronic state
with a fixed number of electrons or spin it is necessary to constrain the search process 
to a particular segment of the qubit Hilbert state.\cite{Jarrod:2016/njp/023023} 
Previously, symmetry constraints were introduced
via penalty functions.\cite{CVQE} 
These penalties guarantee the optimal wavefunction to have the correct symmetry  
only in its expectation values while corresponding variances can be nonzero. 
Recently, solutions to this problem for some symmetries was proposed through introducing 
symmetry-preserving circuits.\cite{Gard:2019vd,Ganzhorn:2018vq} 
Additionally, using symmetries one can reduce the number of qubits in the Hamiltonian 
and thus facilitate optimization of the electronic state.\cite{Bravyi:2017/ArXiv/1701.08213}

Interestingly, symmetries in VQE are not only necessities for accessing higher states in the Fock space
but also means to improve the accuracy. Some simple symmetries (e.g. the parity of the electronic number) 
were used to mitigate errors originating from noise by measuring entangled ancilla qubits.
\cite{PhysRevA.98.062339,McArdle:2019bo} Also, due to commutativity of some Hamiltonian 
terms with the electron number symmetry operator, errors in preparation or measurement of the wavefunction
can be identified from the wavefunction read-out after measurement of these terms.\cite{CVQE}

In this paper, we investigate whether projectors based on Hamiltonian symmetries can be a better alternative to constraints. General procedures  
of introducing both constraints and projectors are considered.  A particular attention is dedicated to 
generating compact qubit forms for symmetry projectors, which was a problem discovered in earlier 
studies.\cite{Moll:2016gx,Ryabinkin:2018tk} This appears to be possible only for 
some symmetries while for others only approximate expressions are feasible. 
Generally, time complexity of a single step of VQE is tied to the number of terms to be measured 
and their variances. Only single qubit measurements are available in the
current architectures, this limits the elementary measurable operator parts to linear combinations
of operators that commute with each other at a single qubit level.\cite{Verteletskyi:2019mcc} 
Recently, several new techniques to reduce the number of linear combinations requiring separate measurements
have been introduced.\cite{Izmaylov:2019gb,Izmaylov:2019ac,Yen:2019c}

Apart from imposing physical symmetries to converge to the right state, 
symmetry projectors also allow more efficient search for the wavefunction of strongly correlated 
systems on a classical computer.\cite{Qiu:2018ei} 
In this context, introducing symmetry projectors into the 
variational procedure ({\it i.e.} variation-after-projection type of methods) 
provides more efficient use of variational parameters in the wavefunction ansatz. 
One rationale for this is that the wavefunction ansatz does not need to use 
variational parameters for maintaining the right symmetry but only to lower the energy.
Thus, development of projection techniques in quantum computing can be an efficient 
approach to treatment of strongly correlated systems.

The rest of the paper is organized as follows. Section~\ref{sec:theory} presents theory of 
symmetry projection construction, compares different ways of use of projectors and constraints in 
the VQE procedure, discusses approaches to building approximate projectors and their relations with 
constraints, and provides qubit space expressions for operators involved in VQE optimization.   
In Sec.~\ref{sec:numer-stud-disc} we assess various symmetry projection and constraint techniques on a set
of three molecular systems (\ce{H2}, \ce{LiH}, and \ce{H2O}) within the qubit mean-field approach.
Section~\ref{sec:conclusions} concludes by providing summary and outlook.     

\section{Theory}
\label{sec:theory}

\subsection{Use of symmetry}

Here we review basic elements of symmetry use in the eigenvalue problem. Most of the material can be  
found in various textbooks,\cite{Barut,Fuchs,Gilmore:2008} but to keep the paper self-contained we review it here with 
a special emphasis on treatment of multiple symmetry operators that do not generally commute with each
other (i.e. non-abelian case).

For any Hamiltonian one can find a set of operators $\{\hat O_i\}$ commuting with the Hamiltonian, 
$[\hat H,\hat O_i] = 0$. However, in general, these operators do not commute with each other $[\hat O_i,\hat O_j]\ne 0$. 
On the one hand, non-commutativity introduces a problem that there is no common set of eigenfunctions for 
all symmetry operators, but on the other hand, it allows us to generate additional operators 
that commute with $\hat H$ by forming all possible nontrivial commutators within the $\{\hat O_i\}$ set. 
This process leads to an algebraic structure, where all $\hat O_i$ operators in the augmented set 
satisfy the condition
\bea
[\hat O_i,\hat O_j] = \sum_k c_{ij}^{(k)} \hat O_k, 
\eea
here $c_{ij}^{(k)}$ are some constants. 
Mathematically speaking, we have obtained a Lie algebra which consists of 
operators commuting with the Hamiltonian and thus are referred to as symmetries. 

In this general case, the eigenfunctions of the Hamiltonian correspond to particular irreducible eigen-subspaces 
of the symmetry operators which are not necessarily one dimensional as in the abelian case. Considering 
the structure of these irreducible eigen-subspaces is necessary to build proper constraints or projectors.  
 
For the purpose of construction projectors on eigen-spaces of the symmetry operators $\{\hat O_i\}$
it is useful to separate two cases based on whether in addition to the Lie algebraic structure 
$\{\hat O_i\}$  form a multiplicative group $\hat O_i\hat O_j = \hat O_k$ (e.g. point group symmetries) 
or not (e.g. electron spin $su(2)$ Lie algebra). 


Existence of the group structure allows {one} to generate projectors on the group irreducible 
representations following the standard procedure
\bea\label{eq:FG}
\hat P_\Gamma = \frac{{d_\Gamma}}{|G|} \sum_{k=1}^{|G|} \chi_\Gamma^{*}(\hat O_k) \hat O_k, 
\eea 
where $\Gamma$ is the irreducible representation of interest, {$d_\Gamma$ is the dimension of $\Gamma$}, and $\chi_{\Gamma}(\hat O_k)$ are characters for 
the group elements. $\hat O_k$ are generally not unitary operators, but any finite group can be represented as 
a set of unitary operators. Therefore, we will consider $\hat O_k$'s forming a finite group as unitary.    

In absence of a group structure, the Lie algebra can be turned into a continuous Lie group using 
the standard exponential mapping. Then the same standard machinery as in finite groups can be extended
to continuous compact groups to formulate projectors.  
However, switching from the algebra to a group is not necessary to obtain the projectors 
onto irreducible representations of the algebra. Moreover, irreducible 
representations of the underlining Lie algebra are still necessary for 
constructing projectors onto irreducible representations of the group. 

Standard techniques to build irreducible representation of 
simple and semisimple Lie algebras (e.g. $su(2)$, the electron spin) are well described in various 
mathematical textbooks\cite{Fuchs,Barut} and Ref.~\onlinecite{Izmaylov:2019bx}. 
We will not detail them here but only formulate the useful result that in all semisimple Lie algebras, 
it is straightforward to find a set of fully commuting operators whose eigenvalues characterize 
and eigenfunctions span all irreducible representations.   
For the well-known $su(2)$-case, the usual choice of these operators are $\hat S_z$ and $\hat S^2$. 
From the computational point of view, it is convenient to present projectors for each 
of the operators in the commuting set $\hat O_i$ as an operator function
$\hat P_j^{(i)} = F(\hat O_i,o_j^{(i)})$, where $o_j^{(i)}$ is the eigenvalue determining the 
eigen-subspace of interest. Then the total projector can be written as 
\bea\label{eq:alg_proj}
\hat P_{\Gamma} = \prod_i  F(\hat O_i,o_j^{(i)}), ~ j\in\Gamma
\eea
where the eigenvalues $o_j^{(i)}$ should be chosen so that the projection is done 
on a particular irreducible subspace of the Lie algebra, $\Gamma$. All operators $\hat O_i$ in 
\eq{eq:alg_proj} commute, and therefore, their projectors can be put in any order. 

\subsection{Construction of projectors for individual symmetry operators}
\label{sec:ex_proj}

For practical use of \eq{eq:alg_proj} we summarize few approaches for constructing functional forms for individual 
symmetry projectors $\hat P_j^{(i)}=F(\hat O_i,o_j^{(i)})$, while more detailed discussion is provided 
in Ref.~\onlinecite{Izmaylov:2019bx}. The majority of symmetry operators have discrete spectra, and 
the corresponding function $F$ can be constructed from  
some differentiable representation of the Kronecker-delta function.
To see this, let us present the projector as 
\bea\label{eq:DK}
\hat P_j^{(i)} &=& \sum_n \ket{\phi_n^{(i)}}\bra{\phi_n^{(i)}} \delta_{nj}\\
&=& \sum_n \ket{\phi_n^{(i)}}\bra{\phi_n^{(i)}} F(x,o_j^{(i)})\vert_{x=o_n^{(i)}},
\eea  
where $\ket{\phi_n^{(i)}}$ are the eigenfunctions of $\hat O_i$ corresponding to eigenvalues $o_n^{(i)}$.  
Here, we substituted the Kronecker-delta function $\delta_{nj}$ with a differentiable function $F(x,o_j^{(i)})$
of the form
\bea\label{eq:gen}
F(x,o_j^{(i)}) =
\begin{cases}
1,~ x=o_j^{(i)}, \\
0,~ x=o_n^{(i)}, n\ne j, \\
\xi(x) \in [0,1],~ x \ne o_n^{(i)}, \forall n,
\end{cases}
\eea
where $\xi(x)$ can be any smooth function for intermediate values of $x$. 
Due to its differentiability, one can expand $F$ in the Taylor series, and this expansion defines 
$F(\hat O_i, o_j^{(i)})$. 
There are multiple ways to define $F(x,o_j^{(i)})$,\cite{Izmaylov:2019bx} 
here we list the two most useful:\\
{\it 1) Integration over a unit circle:} 
\bea\label{eq:exp1}
F(x,o_j^{(i)}) = \frac{1}{2\pi}\int_0^{2\pi} e^{i\phi(x-o_j^{(i)})}d\phi.
\eea
Here, for any $x\ne o_j^{(i)}$ we obtain zero. Such selectivity comes with a price of introducing
the integral. \\
{\it 2) The Lagrange interpolation product:} 
\bea\label{eq:LIP}
F(x,o_j^{(i)}) = \prod_{n\ne j} \frac {x - o_n^{(i)}}{o_j^{(i)}- o_n^{(i)}}, 
\eea
which is less restrictive since for $x$-values in between the eigenvalues the functional value is not fixed 
to zero or one. The L\"owdin spin projector uses \eq{eq:LIP}.\cite{Lowdin}   
 
Equation \eqref{eq:LIP} is especially useful to build projectors for operators with finite number of 
eigenvalues. Interestingly, for such operators, projectors based on Eqs.~\eqref{eq:exp1} and \eqref{eq:LIP} 
are the same. This is a consequence of the Cayley-Hamilton theorem\cite{Gelfand} because 
there is only a finite number of linear independent powers for operators with finite spectra. 
Thus, any function of such an operator is equivalent to $M-1$ polynomial, where $M$ is the number of 
eigenvalues. 
 
Another interesting connection can be found between projectors based on \eq{eq:exp1} and generalization 
of the group projector in \eq{eq:FG} to a continuous group.    
It is straightforward to see that $g(\hat O_i,\phi) = \exp[i\phi \hat O_i]$, where $\phi\in[0,2\pi]$ forms a continuous compact  
one-parametric cyclic group, $g(\hat O_i,\phi')g(\hat O_i,\phi'')=g(\hat O_i,\phi)$, $\phi=\mod(\phi'+\phi'',2\pi)$. 
All cyclic groups are abelian and have one-dimensional irreducible 
representations, in this case, irreducible representations are characterized by $o_j^{(i)}$, with characters 
$\exp[i\phi o_j^{(i)}]$. Therefore,  
\bea\label{eq:IG}
\hat P_j^{(i)} = \frac{1}{2\pi}\int_0^{2\pi} d\phi  e^{i\phi (\hat O_i-o_j^{(i)})},
\eea 
can be interpreted both as the result from \eq{eq:exp1} and 
as the continuous group extension of \eq{eq:FG}.

\subsection{Constraining the energy minimization}

Two main approaches to impose symmetry constraints in the variational search are addition of penalty 
functions and projecting out irrelevant symmetries. 
We give a brief overview of different schemes within these two approaches.
It will be assumed that some unitary parametrization is used for the wavefunction, $\ket{\psi(\theta)}$, where 
$\theta$ is a set of parameters. Note that the only difference from the conventional VQE scheme 
will be in minimization of different functionals involving symmetry operators. 
To evaluate these functionals in the VQE fashion, {\BC a few extra expectation values need to be measured.}


{\it Adding penalty for deviation from correct average values} 
requires minimization of the following functional
\bea\label{eq:CVQE}
E_c[\ket{\psi(\theta)}] &=& \bra{\psi(\theta)} \hat H \ket{\psi(\theta)} \\ \notag
&+& \mu \sum_i (\bra{\psi(\theta)} \hat O_i \ket{\psi(\theta)}-o_j^{(i)})^2,
\eea
where $\{\hat O_i\}$ are mutually commuting symmetry operators with their 
eigenvalues $o_j^{(i)}$ defining a certain irreducible subspace. 
This approach has been implemented in constrained VQE (CVQE) and has 
advantages of simplicity (only averages of the symmetry operators are needed), 
the shortcomings of CVQE is that the symmetry is satisfied only on average, 
and it is possible that variances of $\hat O_i$ are non-zero. 

{\it Adding penalty for deviation from the correct averages and non-zero variances}  
extends the Lagrange functional of \eq{eq:CVQE} by adding penalties for variances
\bea\notag
\tilde{E}_c[\ket{\psi(\theta)}] &=& E_c[\ket{\psi(\theta)}]  + \mu\sum_i (\bra{\psi(\theta)} \hat O_i^2 \ket{\psi(\theta)}\\ \label{eq:CVAR}
&& -\bra{\psi(\theta)} \hat O_i \ket{\psi(\theta)}^2).
\eea  
It is straightforward to show that the variance of the operator reaches 
its minimum (zero) only on its eigenstates (see appendix A).  
Thus, for the computational price of evaluating expectation 
values of squares of the symmetry operators we can impose the symmetry completely. 
Also, minimization in \eq{eq:CVAR} is equivalent to modification of the Hamiltonian as 
\bea
\hat H_c = \hat H + \mu\sum_i (\hat O_i -o_j^{(i)})^2
\eea
in the regular variational procedure.

One unpleasant feature of both constraint approaches is the presence of arbitrary parameter
$\mu$ which is usually set to a large positive number. This arbitrariness does 
not affect results if the variational ansatz $\ket{\psi(\theta)}$ is flexible enough to satisfy the 
constraint exactly. However, if this condition is not satisfied, $\mu$ can 
significantly affect the final energy. 

Another limitation for both approaches is that treatment of non-abelian 
group symmetries will require introducing the projector on irreducible subspace (\eq{eq:FG}).  

{\it Introducing projectors to penalize undesired symmetry components} 
addresses constraining the energy minimization for non-abelian groups.    
If the projector to an eigenstate of the right symmetry is available then we 
can modify the Hamiltonian to introduce the penalty for components of undesired 
symmetry
\bea\label{eq:cHG}
\hat H_\Gamma  = \hat H + \mu (\mathbb{1} - \hat P_{\Gamma}).
\eea
Here we used the idempotency condition for the projector ($\hat P_{\Gamma}^2 = \hat P_{\Gamma}$) so that 
non-negative operator $(\mathbb{1}-\hat P_{\Gamma})^2$ is reduced  to $(\mathbb{1}-\hat P_{\Gamma})$. 
This approach is generally equivalent to imposing constraints on averages and variances, but it has an 
advantage to be able to address the non-abelian group symmetry cases. 
On the other hand, it can introduce more complex dependence on the symmetry operators from \eq{eq:alg_proj} 
for $\hat P_{\Gamma}$.

{\it Projecting out undesired symmetries} introduces 
the functional that projects out undesired symmetry components from the wavefunction
\bea
E_\Gamma[\ket{\psi(\theta)}] &=& \frac{ \bra{\psi(\theta)} \hat P_\Gamma^{\dagger} \hat H \hat P_\Gamma \ket{\psi(\theta)}}{ \bra{\psi(\theta)} \hat P_\Gamma^{\dagger} \hat P_\Gamma \vert\psi(\theta)\rangle}  \\ \label{eq:HPdP}
&=&    \frac{ \bra{\psi(\theta)}  \hat H \hat P_\Gamma \ket{\psi(\theta)}}{ \bra{\psi(\theta)} \hat P_\Gamma \vert\psi(\theta)\rangle},
\eea
where the second equality is a result of the projector's commutation with $\hat H$, 
hermiticity, and idempotency. Note that the point of view that the projector modifies the wavefunction 
makes \eq{eq:HPdP} significantly different from the approaches based on penalties. Indeed,
having the projector in the denominator is an essential feature that differentiates this expression 
from other forms. To see this, let us consider an alternative, where a function commuting with  
the Hamiltonian $\hat\FF$ is introduced only to the numerator 
\bea   
E_{\FF}[\ket{\psi(\theta)}] &=& \bra{\psi(\theta)} \hat\FF^{\dagger} \hat H \hat\FF \ket{\psi(\theta)} \\
\label{eq:ST}
&=& \bra{\psi(\theta)}  \hat H \hat\FF^{\dagger}\hat\FF \ket{\psi(\theta)} \\
&=& \bra{\psi(\theta)}  \hat H  + \mu (\mathbb{1} - \hat P_{\Gamma}) \ket{\psi(\theta)},
\eea
where in the last equality we defined $\hat\FF^{\dagger}\hat\FF = \mathbb{1} + \mu 
\hat H^{-1} (\mathbb{1} - \hat P_{\Gamma})$. 
Thus $\FF$ can be associated with $\hat P_{\Gamma}$ and
variational optimization of $E_\FF[\ket{\psi(\theta)}]$ is equivalent to the optimization 
of the $\hat H_{\Gamma}$ Hamiltonian in \eq{eq:cHG}.

\subsection{Approximate projectors} 

As we will see further the exact projector expressions are not always feasible for 
an efficient implementation. Here we present two approaches for approximating 
projectors. 

\subsubsection{Group theoretical approximation: Forming subgroups}

To simplify general expressions for the group projection in \eq{eq:FG} or in the analogous 
infinite summation over the cyclic group in \eq{eq:IG}, one can reduce the summation to that
over a subgroup that permits an efficient implementation. For example,
integration over an infinite cyclic group can be substituted by a summation over a finite 
cyclic subgroup. Note that for operators with spectra where ratios of spectral gaps for 
the eigenvalue with its neighbors form a finite set of rational numbers 
(e.g., $\hat S^2$, $\hat S_z$ or the number of electrons operator, $\hat N$) 
one can form a finite cyclic subgroup $\{\hat U^k\}$, 
with the generator $\hat U = \exp(2\pi i \hat O_i /(dM))$,
where $dM$ is a scaling factor that makes all eigenvalues of $\hat O_i$ to be rational numbers. 
The characters of sought irreducible representations are 
$\chi_\Gamma (\hat U^k) = \exp(2\pi i \Gamma k /(dM))$, where $\Gamma$'s are eigenvalues of $\hat O_i$, $o_j^{(i)}$. 
The size of $\{\hat U_k\}$ can be regulated by choosing $dM$.  

\subsubsection{Appoximating the Kronecker-delta function}

The exact projector operator was obtained using a form of continuous indicator function 
for the corresponding symmetry operator (\eq{eq:gen}). To introduce approximations 
to projector operator one can be less strict on how many eigen-values will be zeroed 
by an approximate version of the continuous representations of the Kronecker-delta function  
\bea\label{eq:FF}
\FF(x,o_j^{(i)}) =
\begin{cases}
1,~ x=o_j^{(i)}, \\
0,~ x=o_n^{(i)},~ o_n^{(i)}\in \mathbf{S}, \\
\xi(x),~ x \not\in  \mathbf{S} \cup o_j^{(i)}, 
\end{cases}
\eea
where $\xi(x)$ is an arbitrary function maintaining smoothness, and $\mathbf{S}$ is an incomplete set 
of eigen-values for the target symmetry operator that one would like to eliminate. This definition does not 
guarantee the idempotency when the Hermitian symmetry operator $\hat O_i$ is made an argument, 
$\FF(\hat O_i,o_j^{(i)})^2\ne\FF(\hat O_i,o_j^{(i)})$  but still preserves the Hermiticity 
$\FF(\hat O_i,o_j^{(i)})^{\dagger} = \FF(\hat O_i,o_j^{(i)})$. Using this approximation in the variational approach results in
\bea
E_\FF[\psi] &=& \frac{\bra{\psi}\hat\FF^\dagger \hat H \hat\FF \ket{\psi}}{\bra{\psi}\hat\FF^\dagger
\hat\FF \ket{\psi}}\\ \label{eq:Fp}
&=& \frac{\bra{\psi} \hat H \hat\FF^2 \ket{\psi}}{\bra{\psi}\hat\FF^2 \ket{\psi}}.
\eea
The wavefunction in this functional can be expanded in mutual eigenstates $\{\phi_k\}$ of 
$\hat H$ and symmetry operator $\hat O_i$
\bea\label{eq:PA}
E_\FF[\psi] &=& \frac{\sum_k e_k |c_k|^2 f_k^2}{\sum_k |c_k|^2 f_k^2},
\eea
where $\ket{\psi} = \sum_k c_k\ket{\phi_k}$,  $\hat H \ket{\phi_k} = e_k \ket{\phi_k}$, 
and $\hat \FF\ket{\phi_k} = f_k\ket{\phi_k}$. $f_k$ is one for the target symmetry state, 
is zero for $\ket{\phi_k}$ corresponding to symmetries from the elimination 
set $\mathbf{S}$, and is greater than one for all other states. If the target state has the lowest energy 
among all states excluding those from the $\mathbf{S}$ set, the variational procedure will easily converge to the 
target state. Due to variational procedure, the only spurious symmetry 
components in $\ket{\psi}$ after the optimization of \eq{eq:Fp} can be from the $\mathbf{S}$ set. 
Thus, $\ket{\psi}$ can be further purified by application of the $\hat\FF$ operator.

In practice, projectors satisfying these constraints can be constructed through \eq{eq:LIP} 
\bea
  \FF^2(x, o_j^{(i)}) = \prod_{n}^{even} \frac{x - o_n^{(i)}}{o_j^{(i)} - o_n^{(i)}}\prod_{k}^{even}\frac{x - o_k^{(i)}}{o_j^{(i)}- o_k^{(i)}} \label{eq:AP}
\eea  where $\mathbf{S}$ consists of even numbers of $o_k^{(i)}$ and $o_n^{(i)}$ such 
that $o_k^{(i)} < o_j^{(i)} < o_n^{(i)}$. To minimize the number of terms in \eq{eq:AP} one can take 
zero number of terms for one of the sets ($o_k^{(i)}$ or $o_n^{(i)}$) and two for the other.  

\subsection{Operators in the qubit space} 

Here we summarize the Hamiltonian and all of its symmetry operators for molecules in the qubit space. 
For all operators, their fermionic second-quantized form, JW- and BK-transformed qubit forms, exact
and approximate projectors are discussed.

\subsubsection{Hamiltonian}

 In order to formulate the electronic structure problem for a
quantum computer that operates with qubits (two-level systems), the
electronic Hamiltonian needs to be transformed iso-spectrally to its
qubit form. This is done in two steps. First, the
second quantized form of $\hat H_e$ is obtained
\begin{equation}
  \label{eq:qe_ham}
  \hat H_e = \sum_{pq} h_{pq} {\hat a}^\dagger_p {\hat a}_q + \frac{1}{2}\sum_{pqrs}
  g_{pqrs} {\hat a}^\dagger_p {\hat a}^\dagger_q {\hat a}_s {\hat a}_r,
\end{equation}
where ${\hat a}_p^\dagger$ (${\hat a}_p$) are fermionic creation
(annihilation) operators, $h_{pq}$ and $g_{pqrs}$ are one-
and two-electron integrals in a spin-orbital
  basis.\cite{Helgaker:2000} This step has polynomial complexity and
is carried out on a classical computer. Then, using the
\gls{JW}~\cite{Jordan:1928/zphys/631, AspuruGuzik:2005/sci/1704} or
more resource-efficient \gls{BK}
transformation~\cite{Bravyi:2002/aph/210, Seeley:2012/jcp/224109,
  Tranter:2015/ijqc/1431, Setia:2017/ArXiv/1712.00446,
  Havlicek:2017/pra/032332}, the electronic Hamiltonian is converted
iso-spectrally to a qubit form
\begin{equation}
  \label{eq:spin_ham}
  \hat H_q = \sum_I C_I\,\hat W_I,
\end{equation}
where $C_I$ are numerical coefficients, and $\hat W_I$ are
Pauli  ``words", products of Pauli operators of different qubits 
\begin{equation}
  \label{eq:Pi}
  \hat W_I = \cdots \hat \sigma_{2}^{(I)} \, \hat \sigma_{1}^{(I)},
\end{equation}
$\hat \sigma_i^{(I)}$ is one of the $\hat x,\hat y,\hat z$ Pauli
operators for the $i^{\rm th}$ qubit. The number of qubits $N$ is equal 
to the number of spin-orbitals used in the second quantized form [\eq{eq:qe_ham}].
Since every fermionic operator is substituted by a product of Pauli operators in 
both JW and BK transformations, the total number of Pauli words in $\hat H_q$ 
scales as $N^4$. 

\subsubsection{Electron number operator}

The electron number operator has the following forms in various representations
\bea
\hat N &=& \sum_{p=1}^{N_o} \CR{p}\AN{p}, \\
\hat N_{\rm JW} &=& \frac{N_q}{2} - \frac{1}{2}\sum_{k=1}^{N_q} \hat z_k, \\ \label{eq:NBK}
 \hat N_{\rm BK} &=&  \frac{N_q}{2} - \frac{1}{2}\sum_{k=1}^{N_q} \hat z_{\underline{F(k)}},
\eea 
where $N_o$ is the number of orbitals ($N_o=N_q$), 
$F(k)$ is the flip set of qubit $k$, and $\underline{F(k)} = F(k)\ \cup\ k$\cite{Seeley:2012/jcp/224109},  
while $\hat z_{\underline{F(k)}}$ stands for $\hat z$'s applied to all qubits in $\underline{F(k)}$. 
Let us consider the exponential form of the projector to the number of electrons
\bea\label{eq:PN}
\hat P_{N} &=& \frac{1}{2\pi} \int_0^{2\pi} d\phi e^{i\phi (\hat N - N)} \\
&=& \frac{1}{2\pi} \int_0^{2\pi} d\phi e^{-i\phi N} \prod_{k=1}^{N_q} e^{i\phi (c_k\hat W_k)} \\ \label{eq:exp}
&=&  \frac{1}{2\pi} \int_0^{2\pi} d\phi e^{-i\phi N} \prod_{k=1}^{N_q} [\cos(c_k\phi)\mathbb{1} \\ \notag
&&+ i\hat W_k \sin(c_k\phi)], 
\eea  
where $\hat W_k$ are Pauli words constituting $\hat N$ and $c_k$ are 
corresponding coefficients, all $\hat W_k$
are mutually commuting therefore exponent of $\hat N$ is presented as a product of $c_k\hat W_k$ exponents.   
The last integral in \eq{eq:exp} contains $2^{N_q}$ terms, 
hence the scaling of the number of terms in this 
projector is exponential with $N_q$. 
The origin of this problem can be traced to ultra-high precision of the projector in \eq{eq:PN}, 
it separates an $N$-electron component from any other component even if the number 
of electrons in the separated components are different from the desired number by infinitesimal 
amount (e.g. $N\pm10^{-10}$). Clearly, such precision is somewhat excessive, and if we construct a finite 
subgroup built of two elements $\{1,\exp[i\pi \hat N]\}$, projectors on irreducible representations of 
this cyclic subgroup can separate even/odd-electron subspaces.  
The important question is whether such reduction would simplify the form 
of the projector operator? It does for the BK and Parity forms of the $\hat N$ (\eq{eq:NBK})
\bea
\hat P_{e/o}  &=& \frac{1}{2}\left\{\mathbb{1} \pm \exp[i\pi \hat N_{\rm BK}]\right\} \\
&=& \frac{\mathbb{1} \pm \hat z_{N_q}}{2} \label{eq:P_eo}.
\eea 
This is a well-known symmetry in the BK or parity transformations, where 
the last qubit encodes information of the parity of the number of electrons. 

\subsubsection{Electron spin operators}

Using the second-quantized expressions for $S_z$ and its BK and JW transformations
one can write 
\bea
\hat S_z &=& \frac{1}{2} \sum_{p=1}^{N_o/2} \CR{p\alpha}\AN{p\alpha} - \CR{p\beta}\AN{p\beta}\\ 
\hat S_{z,\rm JW} &=& \frac{1}{4}\sum_{p=1}^{N_q/2} - \hat z_{p\alpha} + \hat z_{p\beta}\\
 \hat S_{z,\rm BK} &=&  \frac{1}{4}\sum_{k=1}^{N_q/2} - \hat z_{\underline{F(2k-1)}} + \hat z_{\underline{F(2k)}}\label{eq:SzBK}
\eea
Similarly for the $\hat S_+$ component of $\hat S^2 = \hat S_+ \hat S_- + \hat S_z^2 + \hat S_z$ we can write
\bea
\hat S_+ &=& \sum_{p=1}^{N_o/2} \CR{p\alpha}\AN{p\beta} \\ 
\hat S_{+, \rm JW} &=& \frac{1}{4}\sum_{k=1}^{N_q/2} (\hat x_{2k-1} - i\hat y_{2k-1})
(\hat x_{2k} + i\hat y_{2k}) \label{eq:S2JW}\\
 \hat S_{+, \rm BK} &=&  \frac{1}{4} \sum_{k=1}^{N_q/2}(1-z_{\underline{F(2k)} \setminus 2k-1})(\hat x_{2k-1} - i\hat y_{2k-1}) \label{eq:S2BK}\ \
\eea
while $\hat S_- = \hat S_+^\dagger$ in all forms.
Equations~\eqref{eq:SzBK}, \eqref{eq:S2JW} and \eqref{eq:S2BK} assume the spin-orbital's
ordering ($\alpha$, $\beta$, $\alpha$, $\beta$ ...).  
If we use the exponential function to build projectors for $\hat S_z$ and $\hat S^2$ the same problems as 
in the case of $\hat N$ will appear. Moreover, in the $\hat S^2$ case, the exponent contains non-commuting 
Pauli words, which complicates obtaining final expression as a product of exponents of Pauli words even further. 
On the other hand, projector built from $\hat S_z$ has limited use. It can be used to avoid singlet solution through projecting out $S_z = 0$ but it cannot guarantee singlet solution through projecting out all $S_z \neq 0$. 

It is more natural to build approximations for 
projectors of these spin operators based on the L\"owdin projection based on \eq{eq:LIP}
\bea
\hat P_{S} = \prod_{S_j\ne S}\frac{\hat S^2 - S_j(S_j+1)}{S(S+1)- S_j(S_j+1)}.
\eea
contains potentially a large number of powers of the $\hat S^2$ operator, which increases the computational cost 
of $P_S$. To build approximate functions similar to the discussed $\FF$ in \eq{eq:Fp} we suggest to resort 
to products limited in $S$. The spin eigenvalues included in the product correspond to the $\mathbf{S}$ subset 
of \eq{eq:FF} and their eigenstates are projected out exactly. It is assumed that eigenstates that are not 
projected out are higher in energy and the variational procedure will avoid them. To minimize powers of $\hat S^2$ one 
can approximate $\hat\FF^2$ directly using a limited product 
with an additional requirement of non-negativity. Also, it can be assumed that the projection 
on even and odd number of electrons can always be done easily. For example, to construct 
an approximate non-negative singlet projector within the even number of electrons subspace
that will project out triplet states one can use 
\bea
\hat\FF^2_{S=0} = \left(\frac{2-\hat S^2 }{2}\right)\left(\frac{6-\hat S^2 }{6}\right),
\eea
this projector also eliminates quintet states. 

\subsubsection{Point group symmetry operators}

Assuming $\{\hat O_k\}$ are elements of a finite point group $G$, their reducible matrix representations 
$\mathbf{O}^{(k)}$ in a given set of symmetry adapted orbitals  
$\{\phi_1, \phi_2, \hdots, \phi_{N_o} \}$ have elements $O_{ij}^{(k)} = \bra{\phi_i}\hat O_k \ket{\phi_j}$ 
and possess block diagonal forms.  
Dimensionalities of non-zero blocks are determined by dimensionalities of the corresponding irreducible 
representations (e.g., $A,B,E,T,$ {\it etc}). 
All one-dimensional irreducible representations are given by $O_{ii}^{(k)} = \chi_{\Gamma,i}(\hat O_k) \in \{-1,1\}$, 
where $\chi_{\Gamma,i}(\hat O_k)$ is the character for the irreducible representation $\Gamma$
of the $i^{\rm th}$ orbital under action of $\hat O_k$. 

In second quantization, 
the unitary orbital transformation corresponding to $\hat O_k$ is  
\begin{align}\label{GeneralPGOp}
\hat O_k = \exp(-\hat \kappa), \quad
\hat \kappa = \sum_{ij} \kappa_{ij} a_i^\dagger a_j,
\end{align}
where $\kappa_{ij}$ are elements of anti-Hermitian block-diagonal matrix $\boldsymbol{\kappa} = - \ln(\mathbf{O}^{(k)})$. 
Both $\hat \kappa$ and $\boldsymbol{\kappa}$ differ for 
different group elements $\hat O_k$ and have a dependence on $k$,
but for notational simplicity we keep this dependence implicit. 

For one-dimensional irreducible 
representations $\kappa_{ii} \in \{0,i\pi\}$. 
Hence, for the {\it abelian} groups, where all irreducible representations are one-dimensional, 
the second quantized orbital transformation operator 
$\hat O_k$ is 
\begin{align}
\hat O_k = \prod_{j} \exp(-\kappa_{jj} a_j^\dagger a_j). 
\end{align}
This results in the following forms of the JW, parity, and BK representations 
\begin{align} 
\hat O_{k,JW} & = \prod_{i^*} \hat z_{i^*} \label{PG_JW} \\
\hat O_{k,P} & = \prod_{i^*} \hat z_{i^*} \hat z_{i^*-1} \label{PG_P} \\
\hat O_{k,BK} & = \prod_{i^*} \hat z_{\underline{F(i^*)}} \label{eq:PG_BK},
\end{align}
where $\{\phi_{i^*}\}$ is a subset of the orbitals such that $\chi_{\Gamma({i^*})} = -1 \> \forall \> \phi_{i^*}$. 

For a general case of a non-abelian group some orbitals can correspond to irreducible representations of 
non-unit dimensionality. Due to anti-hermiticity, $\hat \kappa$ can be recast as
\begin{align} \label{second_quant_U}
\hat \kappa = & \sum_{i}\kappa_{ii}a_i^\dagger a_i \nonumber \\ & - \sum_{i < j}\big( \Re(\kappa_{ij})(a_j^\dagger a_i - a_i^\dagger a_j )  -  i\Im(\kappa_{ij})(a_i^\dagger a_j + a_j^\dagger a_i) \big),
\end{align}
from which the qubit-space operator may be obtained by fermion-to-qubit mappings
for the JW and party representations
\begin{widetext}
\begin{align}
\hat \kappa_{JW} & = \sum_{i} \frac{\kappa_{ii}}{2}(\mathbb{1}-\hat z_i) - \frac{i}{2}\sum_{i<j} \Big(\Re(\kappa_{ij})(\hat y_i \hat x_j - \hat x_i \hat y_j) - \Im(\kappa_{ij})(\hat y_i \hat y_j + \hat x_i \hat x_j) \Big) \hat z_{i \leftrightarrow j} \\
\hat \kappa_{P} & = \sum_{i} \frac{\kappa_{ii}}{2}(\mathbb{1}-\hat z_i \hat z_{i-1} ) - \frac{i}{2}\sum_{i < j} \Big( \Re(\kappa_{ij})(\hat z_{i-1}\hat x_i \hat y_{j-1} - \hat y_i \hat x_{j-1} \hat z_j) + \Im(\kappa_{ij})(\hat y_i \hat y_{j-1} + \hat z_{i-1} \hat x_i \hat x_{j-1} \hat z_j) \Big) \hat x_{i \leftrightarrow j-1}, 
\end{align}
\end{widetext}
where $\hat \sigma_{i \leftrightarrow j}$ ($\sigma=x,y,z$) denotes products 
$\hat \sigma_{i+1}...\hat \sigma_{j-1}$. The BK transformed expression has a 
similar structure but in its general form is more complicated to write.

Exponentiation of the off-diagonal elements gives a linear combination of Pauli words 
with an upper bound of $2^{2d-1}$ 
terms per a $d$-dimensional block in $\hat \kappa$. Individual blocks commute, thus the total complexity 
of implementing $\hat O_k$ has upper bound $\prod_i^M 2^{2d_i-1}$ for $M$ $d_i$-dimensional blocks. 
In practice, these estimates are too conservative because there are some cancellations that 
require knowledge of a particular algebra. We illustrate the full process of constructing the projectors for all irreducible representations of the $C_{3v}$ group for a doubly degenerate $E$-type orbital basis
in appendix B. 

When constructing finite point group projectors, one can tailor the level of symmetry employed, 
depending on the subspace of interest. For example, given two irreducible representations $\Gamma$ 
and $\Gamma'$ of group $G$ such that projection by $P_{\Gamma}$ and $P_{\Gamma'}$ 
yield two distinct subspaces, one may consider building the projectors in a proper subgroup $H < G$, 
under condition that $\Gamma$ and $\Gamma'$ remain distinct through the \emph{descent} in 
symmetry $H \leftarrow G$. This has the practical advantage of reducing the number of unitary 
operations in $P_{\Gamma}$ and $P_{\Gamma'}$ since $|H| < |G|$. Furthermore, for high symmetry 
molecular systems such as those belonging to linear groups $D_{\infty h}$ and $C_{\infty v}$, 
construction of the projectors in an overgroup $F > G$ may further split the spectrum, such that $G$ 
is a proper subgroup of the highest-order group the polyatomic system of interest belongs to (e.g. $D_{\infty h}$ or $C_{\infty v}$), under condition that irreducible representation $\Gamma$ of group $G$ may split to irreducible representations $\Gamma'$, $\Gamma''$, $\hdots$ by \emph{ascent} in symmetry $G \to F$. 
Thus, the full set of available symmetry elements to the molecular system may be viewed as a practical resource in the context of point group projectors, for the extent of which we employ is available as choice.

\subsubsection{Low-qubit-number symmetries}  
One of the difficulties in implementing projectors of regular symmetries (e.g., number of electrons and spin)
stems from involvement of all qubits in their operators. Here, we suggest that in some cases, 
it is possible to obtain few-qubit operators that commute with the molecular Hamiltonian. 
Finding such symmetry operators can be done by considering zero commutator problem 
$[\hat H,\hat O(\alpha_i)]=0$, where $\alpha_i$ are $\hat O$'s 
parameters as a linear algebra problem in the space of Pauli words. For example, any single 
qubit operator can be parametrized as $\alpha_1 \hat x+ \alpha_2 \hat y+ \alpha_3 \hat z$, its 
commutator equation with the Hamiltonian will have $\alpha_1=\alpha_2=0,~\alpha_3=1$ solution for 
the last qubit in the BK transformed Hamiltonian. Similarly, one can do excessive search with two-qubit
operators where the total number of parameters is 15 for each pair. Building a projector on 
eigen-spaces of such few-qubit operators can be done using the exponentiation (\eq{eq:IG}).    


\section{Numerical results and discussion}
\label{sec:numer-stud-disc}

To assess developed projector expressions we apply them in evaluation 
of \glspl{PES} for the \ce{H2} and \ce{LiH} molecules within the STO-3G basis and for the 
\ce{H_2O} molecule within the 6-31G basis.
To generate qubit Hamiltonians, the BK transformation was used for \ce{H_2} and \ce{H2O}  
while the parity transformation was employed for \ce{LiH}.
For each system two qubits are stationary ($2$nd and $4$th in \ce{H2},
$3$rd and $6$th in \ce{LiH}, and $4$th and $8$th in \ce{H2O}).\cite{QCC} 
Therefore, the $\hat z$ operators for these qubits were substituted with eigenvalues $\pm 1$ 
so that solutions of interest are within the reduced subspace. 
This reduction is equivalent to projecting to the even number of electrons using \eq{eq:P_eo}, 
and thus, the $\hat P_{e/o}$ projector is not going to be used further.  

 We used the qubit mean-field (QMF)\cite{QMF} and qubit coupled cluster 
(QCC)\cite{QCC} wavefunction ans{\"a}tze  for all calculations with the exception of exact energies, 
which were evaluated via full diagonalization of the qubit Hamiltonians.
The QMF wavefunction is a product of single-qubit coherent states,
\bea
  \ket{\mathbf{\Omega}} &=&  \prod_{i=1}^{N_q} \ket{\Omega_i}, \\
  \ket{\Omega_i} &=& \cos\left(\frac{\theta_i}{2}\right)\ket{0} + \sin\left(\frac{\theta_i}{2}\right)e^{i\phi_i}\ket{1},
\eea where $N_q$ is the number of qubits,  $\boldsymbol{\theta} = \{\theta_1, ..., \theta_N\}$ 
and $\boldsymbol{\phi} = \{\phi_1, ..., \phi_N\}$ are the corresponding Bloch angles taken as variational parameters. 
The QCC wavefunction takes the form 
\bea
    \ket{\Psi} = \prod_{k=1}^{N_E} e^{i\tau_k\hat W_k} \ket{\mathbf{\Omega}}
\eea 
where $\tau_k$ are real-valued amplitudes as 
additional variational parameters {\BC, and $\hat W_k$ are the Pauli words [\eq{eq:Pi}] chosen 
using the energy gradient criterion described in Ref.~\citenum{QCC}.}

For the projector formalism, the energies are evaluated through Eqs.~\eqref{eq:HPdP} and \eqref{eq:Fp} for 
exact and approximate projectors respectively. 
Table~\ref{tab:num_term} illustrates the increase in the number of Pauli 
words from introducing projector operators.
To compare the projector formalism with its constraining alternative, we provide results 
of constrained QMF (CQMF) and QCC (CQCC) calculations where both
averages and variances of symmetry operators were constrained [\eq{eq:CVAR}].

\begin{figure}[h!]%
  \centering
  \includegraphics[width=1\columnwidth]{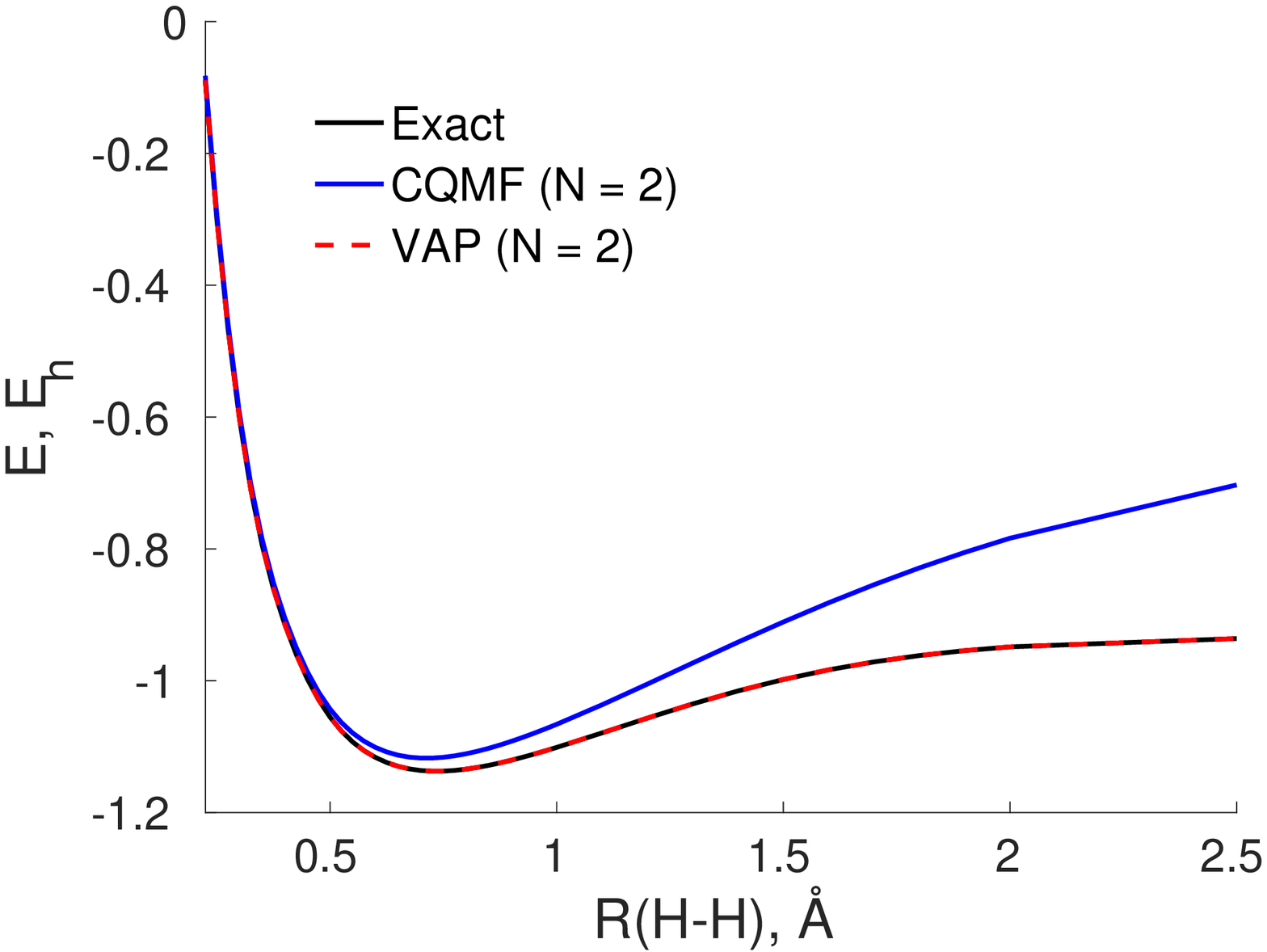}
  \caption{PESs of \ce{H2} evaluated in the neutral ($N = 2$) subspace.} 
  \label{fig:h2_n2}
\end{figure}

\begin{figure}[h!]%
  \includegraphics[width=1\columnwidth]{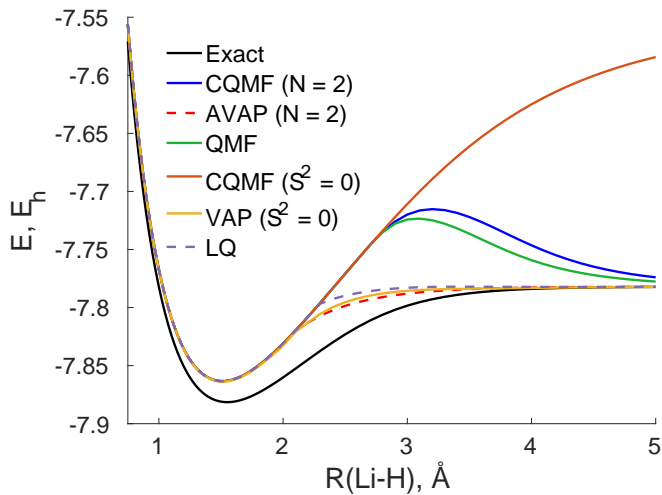}
  \caption{PESs of \ce{LiH} evaluated in variety of subspaces: neutral ($N = 2$), singlet ($S^2 = 0$), 
  and the low-qubit-number (LQ) symmetry with $N + 2S_z = 2$ ($S_z=0$).}
  \label{fig:lih_n2}
\end{figure}

\begin{figure}[h!]%
  \includegraphics[width=1\columnwidth]{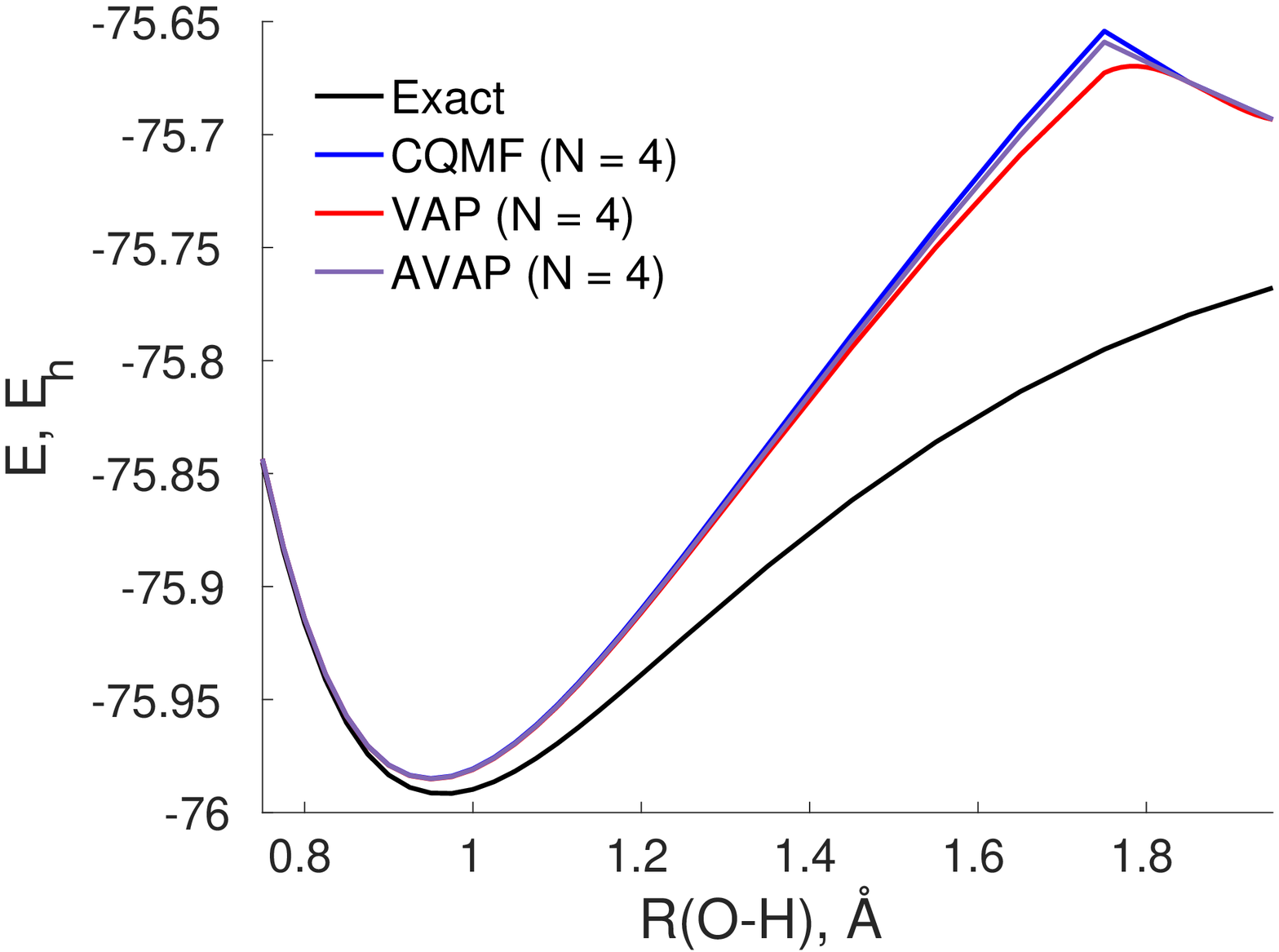}
  \caption{PESs of \ce{H2O} evaluated in the neutral ($N = 4$) subspace. }
  \label{fig:h2o_n4}
\end{figure}

{\it a) Number of electrons:} 
The qubit reduction restricted the number of electrons in the three molecules to the following sets: $(0,\mathbf{2},4)$ 
for \ce{H2}, $(0,\mathbf{2},4,6)$ for \ce{LiH}, and $(0,2,\mathbf{4},6,8)$ for \ce{H2O}, where we highlight in bold the 
neutral configurations. Based on these configurations it is clear that approximation for the electron 
number projector $\FF^2_N$ as in \eq{eq:AP} is only possible for \ce{LiH} and \ce{H2O} by projecting $(4,6)$ and 
$(6,8)$ subspaces respectively.  For \ce{H2}, the only electron number projector is the exact one, $\hat P_{N}$, 
which projects $(0,4)$ subspaces. Variation-after-projection (VAP) with $\hat P_{N}$ for  \ce{H2} recovers the
ground state obtained through full diagonalization of the qubit Hamiltonian (Fig.~\ref{fig:h2_n2}). 
Similarly for \ce{LiH}, $\hat  P_{N}$ in VAP achieves significant energy lowering compared to the constrained 
counterpart. In addition, we found that using approximate projector $\FF^2_N$ results in an identical curve to 
that of $\hat  P_N$ (Fig.~\ref{fig:lih_n2}). The ``hump" on the CQMF ($N = 2$) curve is associated with the spin symmetry 
breaking between singlet and triplet configurations.    
For \ce{H2O}, $\hat P_N$ and $\FF^2_N$ produce insignificant energy lowerings for $R \leq 1.75$\AA, 
for larger bond distances QMF solutions switch to $S^2 = 6$ and exact and approximate projectors do not
affect energy values (Fig.~\ref{fig:h2o_n4}). 

For all three systems projectors result in no more than twice of the number of terms of the original Hamiltonian. Interestingly, due to some term cancellation, the approximate projector generated more terms than 
the exact one in its product with the \ce{H2O} Hamiltonian (see Table~\ref{tab:num_term}). 

\begin{table}[h]%
\setlength\tabcolsep{0pt}
\caption{The number of Pauli words with coefficients $>10^{-9}$ in magnitude for various qubit space operators. 
Subscripts of $\hat P$ and $\hat \FF^2$ denote targeted symmetries, where $LQ$ refers to the low-qubit-number 
symmetries, and $N$ indicates neutral species. 
In rows of two operators the number of unique words is shown. The Hamiltonians are taken at $R = 1$\AA, 
$R = 3.2$\AA~ and $R = 1.95 $\AA~ for \ce{H2}, \ce{LiH} and \ce{H2O}, respectively.}
{
\begin{tabular*}{\columnwidth}{@{\extracolsep{\fill}} l l l l l }
  \toprule
  Operators & \ce{H2} & \ce{LiH} & \ce{H2O} \\
  \midrule
  $\hat H$ & $6$ & $100$ & $165$ \\
  $\hat N$ & $3$  &$7$ & $9$ \\
  $\hat S^2$ & $0$ &$40$ & $71$ \\
  $\hat P_N$ & $2$ &$16$ & $32$ \\
  $\hat H\hat P_N,\ \hat P_N$ & $6$ & $100$ & $264$\\
  $\FF^2_N$ & - &$16$ & $31$ \\
  $\hat H \FF^2_N,\ \FF^2_N$ & - & $100$& $320$ \\
  $\hat H \hat P_{S^2 = 0},\ \hat P_{S^2 = 0}$ &- & $136$& $864$\\
  $\hat H \hat P_{LQ},\ \hat P_{LQ}$ & - & $100$ & $219$ \\
  $\hat P_{B_1}$ &- & - & $2$ \\
  $\hat H \hat P_{B_1}$ &- & - & $242$ \\
  $\hat H \hat P_{B_1} \hat P_{S^2 = 2},\ \hat P_{B_1} \hat P_{S^2 = 2}$ & - & - & $720$ \\
  \bottomrule
\end{tabular*} \label{tab:num_term}
}
\end{table}

\begin{figure}
  \centering
  \includegraphics[width=1\columnwidth]{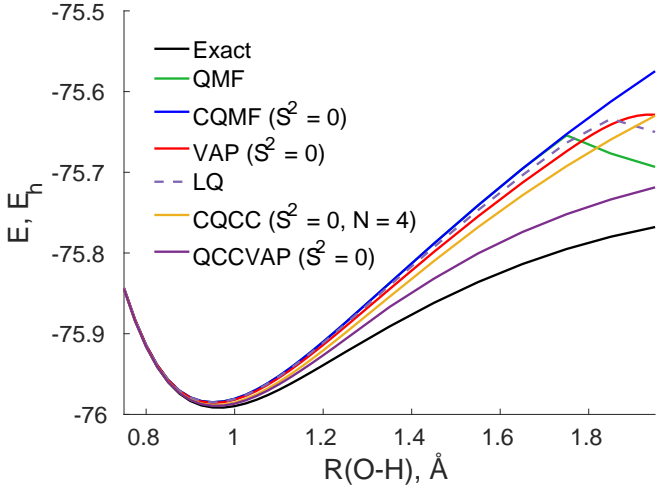}
  \caption{PESs of \ce{H2O} evaluated in the singlet ($S^2 = 0$) subspace and  the low-qubit-number (LQ) symmetry subspace with $N + 2S_z = 4$ ($S_z=0$). Both QCC solutions use 5 entanglers ($\hat x_5 \hat x_4\hat x_3 \hat x_2 \hat x_1 \hat y_0,\ \hat x_4\hat x_3\hat x_1 \hat y_0,\ \hat y_2 \hat x_0,\ \hat x_4 \hat y_1$, and $\hat x_3\hat y_5$).}
  \label{fig:h2o_s0}
\end{figure}

{ \it b) Electron spin:} After the qubit reduction, the \ce{H2} Hamiltonian is in the singlet subspace and therefore does 
not require any spin projection, the \ce{LiH} Hamiltonian has only singlet and triplet subspaces, and the \ce{H2O}
Hamiltonian contains singlet, triplet, and quintet subspaces. These restrictions make approximate singlet 
spin projections equal to the exact one for \ce{LiH} and \ce{H2O}. 
For both systems, the QMF state undergoes symmetry-breaking transition as internuclear 
distance increases. Using the singlet projectors, we obtain singlet neutral solutions for \ce{LiH} 
and \ce{H2O} of lower energies in comparison to CQMF (Figs.~\ref{fig:lih_n2} and \ref{fig:h2o_s0}). 
However, for \ce{H2O}, the QMF quintet ($S^2 = 6$) solutions at $R \geq 1.85$\AA~ are closer to the exact 
ground state energy than the symmetry projected singlet solutions, which indicates need for correlation 
for further improvement of energy of the singlet state. The low energy of the quintet is not accidental because 
for this high spin configuration within the considered qubit space, QMF provides the exact answer. 

\begin{figure}
  \centering
  \includegraphics[width=1\columnwidth]{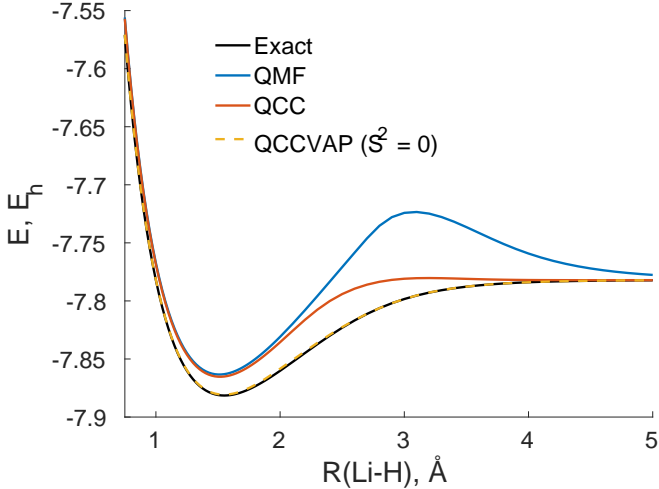}
  \caption{ PESs of \ce{LiH} evaluated using both QMF and QCC wavefunction ans{\"a}tze.
  Both QCC and QCCVAP methods use only the $\hat x_2 \hat x_1 \hat y_0$ entangler.}
  \label{fig:lih_qcc}
\end{figure}

To assess the performance of the singlet projector in correlated methods we employ the 
QCC wavefunction ansatz. 
For \ce{LiH}, to reach $\leq 1\ \text{kcal/mol}$ deviation from the exact result for the whole potential 
energy curve the QCC approach required 7 entanglers. \cite{QCC} 
With the singlet projector, use of only one entangler in QCC, $\hat x_2 \hat x_1 \hat y_0$, 
achieves the same accuracy (Fig.~\ref{fig:lih_qcc}). 
For \ce{H2O}, we employed 5 entanglers with highest gradients found in Ref.~\onlinecite{QCC}: 
$\hat x_5 \hat x_4\hat x_3 \hat x_2 \hat x_1 \hat y_0,\ \hat x_4\hat x_3\hat x_1 \hat y_0,\ \hat y_2 \hat x_0,\ \hat x_4 \hat y_1$, and $\hat x_3\hat y_5$.
Both QCC and QCCVAP curves achieve $\leq 1\ \text{kcal/mol}$ deviation from the exact result 
near the equilibrium geometry, but away from the equilibrium, QCCVAP solutions 
increasingly outperform CQCC solutions (Fig.~\ref{fig:h2o_s0}).


{\it c) Low-qubit-number symmetries:}
For \ce{LiH} and \ce{H2O}, operators involving a half of all qubits are found as the following linear combination 
$\hat O = \hat N + 2\hat S_z$. Projectors targeting the $\hat O$ subspace corresponding to 
neutral closed-shell ($S_z = 0$) species were constructed. 
For \ce{LiH}, the projector achieves significant energy lowering with 
respect to the QMF solution (Fig.~\ref{fig:lih_n2}). In the \ce{H2O} case, minor advantage of the $\hat O$ projector 
is present before symmetry breaking takes place in the QMF solution (Fig.~\ref{fig:h2o_s0}). 
As expected, projectors built from symmetry involving less qubits result in the least overhead of all symmetries 
considered (Table.~\ref{tab:num_term}).

\begin{figure}[h!]
  \centering
  \includegraphics[width=1\columnwidth]{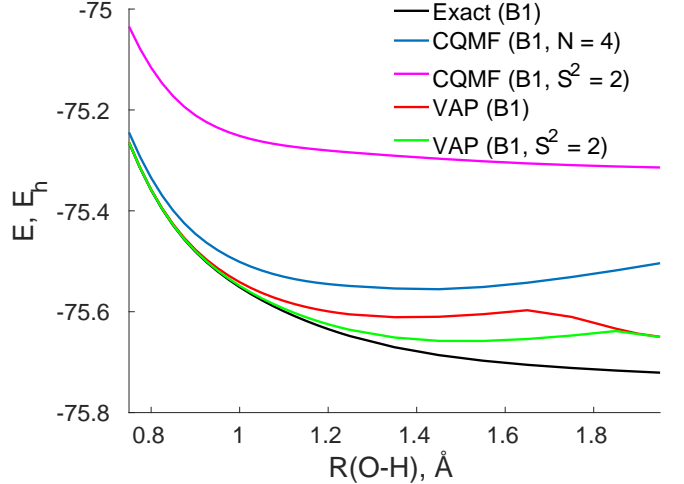}
  \caption{PESs of \ce{H2O} evaluated in the $B_1$ subspace. CQMF constrained to the neutral subspace ($N = 4$) has $S^2 = 1$ whereas CQMF constrained to the triplet ($S^2 = 2$) subspace has $N = 6$. 
  Both solutions of VAP are exactly triplet and neutral on average.}
  \label{fig:h2o_pg_e}
\end{figure}

{ \it d) Point group symmetry: } The reduced qubit Hamiltonians of \ce{H2} and \ce{LiH} are entirely in single 
irreducible representation of the corresponding maximally abelian $D_{2h}$ and $C_{2v}$ groups and thus 
cannot benefit from projection. On the other hand, the \ce{H2O} reduced Hamiltonian contains two irreducible representations of the  $C_{2v}$ group: $A_1$ and $B_1$. The projector on the $A_1$ irreducible subspace 
does not add anything to what was obtained using the number of electrons and spin projectors, therefore
we illustrate capabilities of the projection onto the $B_1$ irreducible subspace. 
The lowest $B_1$ state in the exact solution has also triplet spin symmetry and contains 4 electrons.
If we impose all three symmetry constraints in CQMF, the method cannot converge to a solution. 
This reveals limitations of the QMF wavefunction ansatz that cannot satisfy all constraints. Only two 
out of three symmetries can be satisfied in CQMF (Fig.~\ref{fig:h2o_pg_e}).  
In contrast, the VAP approach can satisfy all symmetry constraints because 
introducing the $B_1$ projection does not require the QMF ansatz to satisfy the point group symmetry. 
Moreover, following this basic idea, 
introducing more than one projector ($B_1$ and $S^2$) lowers the VAP energy even more.  

\section{Conclusions}
\label{sec:conclusions}

We have considered projectors to irreducible subspaces of symmetry operators as alternatives
to symmetry constraints in the variational quantum eigensolver approach to the electronic structure 
problem. Generally, the projector formalism generates a larger number of terms to consider than the constraint 
approach and thus is computationally more expensive. In cases of continuous symmetries 
(e.g. number of electrons and electron spin) the exact projection can involve exponentially large number 
of terms and therefore is infeasible without introducing approximations. Two main approaches to generating 
such approximations have been discussed. A higher number of terms for the projection 
formalism can be intuitively understood considering that the number of terms appearing from 
the multiplication of the Hamiltonian with the symmetry projection is usually larger than that from 
the addition of the constraint to the Hamiltonian. 


The main advantage of using projectors is imposing symmetries without taking resources from 
the variational wavefunction ansatz. This helps to resolve the symmetry dilemma of variational 
ans{\"a}tze : either to lower the energy and break symmetry or to preserve the symmetry but with 
higher energy. In quantum computing, use of projectors can help to reduce the depth of the 
wavefunction generating circuits by shifting the computational burden on the measurement of 
larger number of terms appearing from projection. Additionally, the constraint and projection 
techniques are not mutually excluding and can be used together balancing the number of 
operator terms by placing some 
symmetries as constraints and others as projectors.  

\section{Acknowledgement}
A.F.I. is grateful to I. G. Ryabinkin for useful discussions.  
A.F.I. acknowledges financial support from the Natural Sciences and
Engineering Research Council of Canada and the Ontario 
Early Researcher Award. 


%

\appendix

\section*{Appendix A: Constraining variance imposes the exact symmetry}

Let us consider variance for wavefunction $\ket{\psi}$, which has an orthogonal complement spanned by 
the orthonormal functions $\{\ket{\phi_k}\}$
so that $\ket{\psi}\bra{\psi} + \sum_k\ket{\phi_k}\bra{\phi_k} = \mathbb{1}$ 
\bea
&&\bra{\psi} \hat O_i^2 \ket{\psi}-\bra{\psi} \hat O_i \ket{\psi}^2 = 
\bra{\psi} \hat O_i\mathbb{1}\hat O_i \ket{\psi} \\ \notag
&-&\bra{\psi} \hat O_i \ket{\psi}\bra{\psi} \hat O_i \ket{\psi} \\
&=& \sum_k \bra{\psi} \hat O_i \ket{\phi_k}\bra{\phi_k} \hat O_i \ket{\psi}  \\ \notag
&=& \sum_k |\bra{\psi} \hat O_i \ket{\phi_k}|^2.
\eea
Action of $\hat O_i$ on $\bra{\psi}$ can always be presented as 
\bea\label{eq:O}
\bra{\psi} \hat O_i  = \alpha \bra{\psi}+ \sum_k \beta_k \bra{\phi_k}.
\eea
Using this in the variance expression, we can write
\bea
\sum_k |\bra{\psi} \hat O_i \ket{\phi_k}|^2 &=& \sum_{k} |\sum_{k'} \beta_{k'} \langle \phi_{k'} \ket{\phi_k}|^2 \\
&=& \sum_{k} |\sum_{k'} \beta_{k'} \delta_{kk'}|^2 = \sum_{k} | \beta_{k} |^2,
\eea
therefore, for the zero variance, all $\beta_k$ must be zero, which would make $\bra{\psi}$ an eigenstate 
of $\hat O_i$ according to \eq{eq:O}.

\section*{Appendix B: C$_{3v}$ group projectors for an $E$-type orbital basis}

To illustrate a non-abelian case for point group projection, we consider construction of the projector for irreducible
 representations of $C_{3v}$. Orbitals in a symmetry adapted $C_{3v}$ basis transforming as $A_1$ or $A_2$ 
 do not undergo mixing, and hence corresponding operators are similar to the abelian case. 
 We thus concentrate on a doubly degenerate orbital subset 
 transforming as the $E$-type components $\{ \phi_{x}, \phi_{y} \}$, and assume $\phi_x$ and 
$\phi_y$ are mapped to even index qubits $i$ and $j$ using the BK transformation. We emphasize that the following 
operators are valid only within the degenerate $E$ orbital subspace. The size of the unitary operators required in the 
projector scale by a maximum factor of $2$ for every $A_1$ or $A_2$ adapted orbital added to the basis.
We set $\phi_{x}$ and $\phi_{y}$ such that they transform under action of $\hat C_3$ as
\bea\label{Eorientation}
\hat C_3 \phi_{x} &=& -\frac{1}{2} \phi_x + \frac{\sqrt{3}}{2}\phi_y, \\
\label{Eorientation2}
\hat C_3 \phi_{y} &=& - \frac{\sqrt{3}}{2}\phi_x - \frac{1}{2} \phi_y. 
\eea
The elements of $C_{3v}$ are $\{\mathbb{1}, \hat C_3, \hat C_3^2, \hat \sigma_v, \hat \sigma_v', \hat \sigma_v''\}$, 
thus the form of the projector for an arbitrary irreducible representation 
$\Gamma$ of $C_{3v}$ is 
\bea \label{C_3vprojector}
\hat P_\Gamma &=& \frac{1}{6} \Big{(} \chi_\Gamma^{*}(\mathbb{1}) \mathbb{1} + \chi_\Gamma^{*}(\hat C_3) \hat C_3 + \chi_\Gamma^{*}(\hat C_3^2) \hat C_3^2 \nonumber \\  
&+& \chi_\Gamma^{*}(\hat \sigma_v) \hat \sigma_v + \chi_\Gamma^{*}(\hat \sigma_v') \hat \sigma_v' + \chi_\Gamma^{*}(\hat \sigma_v'') \hat \sigma_v'' \Big{)}.
\eea 
The setting of $\phi_x$ and $\phi_y$ in Eqs. (\ref{Eorientation}) and (\ref{Eorientation2}) causes transformation matrix $\bf{C_3}$ to result in $\kappa_{11} = \kappa_{22} = 0$, $\kappa_{12} =  \frac{2\pi}{3}$ and 
$\kappa_{21} = - \kappa_{12}$. The exponentiated sum for $\hat C_3$ thus takes the form
\begin{align} \label{eq:U^C3}
\hat C_3 & = \exp\big(i\frac{\pi}{3}(\hat x_i \hat x_j + \hat y_i \hat y_j) \hat x_{U_{ij}/\alpha_{ij}} \hat z_{P_{ij}/\alpha_{ij}} \hat y_{\alpha_{ij}} \big) \nonumber \\
& = \frac{1}{4} \mathbb{1} + \frac{3}{4} \hat  z_i \hat z_j + i \frac{\sqrt{3}}{4}(\hat y_i \hat y_j + \hat x_i \hat x_j) \hat x_{U_{ij}/\alpha_{ij}} \hat z_{P_{ij}/\alpha_{ij}} \hat y_{\alpha_{ij}},
\end{align}
where 
\bea
U_{ij} &\equiv& U(i) \bigtriangleup U(j), \\
P_{ij} &\equiv& P(i) \bigtriangleup P(j), \\
\alpha_{ij} &\equiv& U(i) \cap P(j).
\eea
The symbol $\bigtriangleup$ denotes the disjoint union, and $U(i)$ and $P(i)$ respectively denote the update set and parity set of $i$. 

$\hat C_3^2$ is analogous to $\hat C_3$ case but with a sign swapping between $\kappa_{12}$ 
and $\kappa_{21}$, resulting in
\begin{align} \label{U^C3^2}
\hat C_3^2 & = \frac{1}{4} \mathbb{1} + \frac{3}{4} \hat  z_i \hat z_j - i \frac{\sqrt{3}}{4}(\hat y_i \hat y_j + \hat x_i \hat x_j) \hat x_{U_{ij}/\alpha_{ij}} \hat z_{P_{ij}/\alpha_{ij}} \hat y_{\alpha_{ij}}
\end{align}
Setting $\hat \sigma_v$ as the reflection plane defined by the $x$ and $z$ axis, then $\hat \sigma_v \phi_x = \phi_x$ and $\hat \sigma_v \phi_y = - \phi_y$. The only nonzero element in $\bf{\kappa}$ is $\kappa_{22} = i \pi$. 
It follows that $\hat \sigma_v$ takes a simple abelian form
\begin{align} \label{U^sigmav}
\hat \sigma_v = \hat z_{\underline{F(j)}} = \hat z_{j},
\end{align}
where the second equality holds since $j$ was chosen to be even. Setting $\hat \sigma_v'$ as the reflection plane acting on the $E$ components as 
\bea
\hat \sigma_v ' \phi_x &=& - \frac{1}{2} \phi_x + \frac{\sqrt{3}}{2} \phi_y \\
\hat \sigma_v ' \phi_y &=& \frac{\sqrt{3}}{2}\phi_x + \frac{1}{2}\phi_y, 
\eea
gives elements $\kappa_{11} = i3\pi/4$, $\kappa_{22} = i\pi/4$, and $\kappa_{12} = \kappa_{21} = -i\sqrt{3}/4$. 
The resulting qubit space unitary operator is
\begin{align} \label{U^sigmav'}
\hat \sigma_v = & \exp\big(\frac{i\pi}{8}(3\hat z_i + \hat z_j - 4\mathbb{1} \nonumber \\ & -\sqrt{3}(\hat x_i \hat y_j - \hat y_i \hat x_j) \hat x_{U_{ij}/\alpha_{ij}} \hat z_{P_{ij}/\alpha_{ij}} \hat y_{\alpha_{ij}} \big) \nonumber \\
= & \frac{3}{4} \hat z_i + \frac{1}{4} \hat z_j - \frac{\sqrt{3}}{4}(\hat x_i \hat y_j - \hat y_i \hat x_j ) \hat x_{U_{ij}/\alpha_{ij}} \hat z_{P_{ij}/\alpha_{ij}} \hat y_{\alpha_{ij}}.
\end{align}
The remaining reflection plane $\hat \sigma_v ''$ must act on the $E$ components as 
\bea
\hat \sigma_v '' \phi_x &=& - \frac{1}{2} \phi_x - \frac{\sqrt{3}}{2}\phi_y,\\
\hat \sigma_v '' \phi_y &=& - \frac{\sqrt{3}}{2} \phi_x + \frac{1}{2}\phi_x \phi_y, 
\eea
\\
\noindent leading to same case as $\hat \sigma_v '$ but with positive $\kappa_{12}$ and $\kappa_{21}$, giving qubit unitary form:
\bea \notag
\hat \sigma_v'' &=& \frac{3}{4} \hat z_i + \frac{1}{4} \hat z_i \\ \label{U^sigmav''}
&+& \frac{\sqrt{3}}{4}(\hat x_i \hat y_j - \hat y_i \hat x_j ) \hat x_{U_{ij}/\alpha_{ij}} \hat z_{P_{ij}/\alpha_{ij}} \hat y_{\alpha_{ij}}.
\eea
Further simplification may arise by classes present within the point group of interest. The two non-trivial classes in $C_{3v}$ are $\{\hat C_3, \hat C_3^2 \}$ and $\{\hat \sigma_v , \hat \sigma_v ',\hat \sigma_v ''\}$. Since elements within a class will have identical characters for a given irreducible representation, unitaries in the projector expression may be factored leading to $n$ groupings of unitary operations for $n$ classes within the group, leading to favorable cancellation of multi-qubit operations in Eq.(\ref{C_3vprojector}). The resultant $C_{3v}$ projector is
\begin{align}
\hat P_\Gamma = \frac{d_\Gamma}{6} \big(  & \chi_\Gamma^{*}(\mathbb{1}) \mathbb{1} + \frac{\chi_\Gamma^{*}(\hat C_3)}{2} (\mathbb{1} + 3 \hat z_i \hat z_j) + \frac{3\chi_\Gamma^{*}(\hat \sigma_v)}{2} (\hat z_i + \hat z_j) \big).
\end{align}  

\end{document}